\documentclass[sigconf,nonacm]{acmart}

\usepackage{todonotes}
\usepackage{tcolorbox}

\tcbuselibrary{breakable}

\AtBeginDocument{%
  \providecommand\BibTeX{{%
    \normalfont B\kern-0.5em{\scshape i\kern-0.25em b}\kern-0.8em\TeX}}}

\setcopyright{acmcopyright}
\copyrightyear{2018}
\acmYear{2018}
\acmDOI{XXXXXXX.XXXXXXX}

\acmConference[Conference acronym 'XX]{Make sure to enter the correct
  conference title from your rights confirmation emai}{June 03--05,
  2018}{Woodstock, NY}
\acmPrice{15.00}
\acmISBN{978-1-4503-XXXX-X/18/06}

\begin{document}

\title{Experiences from Integrating Large~Language~Model~Chatbots~into~the~Classroom}

\author{Arto Hellas}
\affiliation{%
  \institution{Aalto University}
  \city{Espoo}
  \country{Finland}
}
\email{arto.hellas@aalto.fi}
\orcid{0000-0001-6502-209X}

\author{Juho Leinonen}
\affiliation{%
  \institution{Aalto University}
  \city{Espoo}
  \country{Finland}
}
\email{juho.2.leinonen@aalto.fi}
\orcid{0000-0001-6829-9449}

\author{Leo Leppänen}
\affiliation{%
  \institution{University of Helsinki}
  \city{Helsinki}
  \country{Finland}
}
\email{leo.leppanen@helsinki.fi}
\orcid{0000-0003-3969-8410}


\begin{abstract}
  In the present study, we provided students an unfiltered access to a state-of-the-art large language model (LLM) chatbot. The chatbot was intentionally designed to mimic proprietary commercial chatbots such as ChatGPT where the chatbot has not been tailored for the educational context; the underlying engine was OpenAI GPT-4. The chatbot was integrated into online learning materials of three courses. One of the courses focused on software engineering with LLMs, while the two other courses were not directly related to LLMs.  Our results suggest that only a minority of students engage with the chatbot in the courses that do not relate to LLMs. At the same time, unsurprisingly, nearly all students in the LLM-focused course leveraged the chatbot. In all courses, the majority of the LLM usage came from a few superusers, whereas the majority of the students did not heavily use the chatbot even though it was readily available and effectively provided a free access to the OpenAI GPT-4 model. We also observe that in addition to students using the chatbot for course-specific purposes, many use the chatbot for their own purposes. These results suggest that the worst fears of educators -- all students overrelying on LLMs -- did not materialize even when the chatbot access was unfiltered. We finally discuss potential reasons for the low usage,  suggesting the need for more tailored and scaffolded LLM experiences targeted for specific types of student use cases.
\end{abstract}

\ccsdesc[500]{Social and professional topics~Computing education}

\keywords{large language models, chatbots, classroom experiences, usage analysis, experience report}


\maketitle

\section{Introduction}

Large Language Models (LLMs) such as ChatGPT have captured the attention of both the academia and the general public. Initial hype---especially outside of academic works---has framed LLMs as omnipotent replacements for every creative and knowledge worker. Reactions to the use of LLM-based generative systems in academic settings have been mixed, ranging from calls for---and realized---bans~\cite{de2023top} to claims that they are the new normal and teaching should be reorganized around them~\cite{mehta2023banning}.

The avalanche of LLMs is visible also in computing education and computing education research, where researchers have highlighted a variety of tasks that LLMs can do~\cite{denny2023computing,prather2023robots}. As students begin to use these tools, several of the threats identified by researchers have come into clearer focus. For instance, students often do not understand the code automatically generated by LLMs because they did not write it~\cite{prather2023its}. Even so, students may quickly accept incorrect code suggestions and tinker with the code before discovering they do not need it, only to start over again~\cite{prather2023its,kazemitabaar2023studying, vaithilingam2022expectation}. More generally, there's evidence that the use of LLM assistants may, for example, lead to programmers writing less secure code~\cite{perry2023users}.

In this article, we outline our experiences from integrating a state-of-the-art LLM-powered\footnote{We used GPT-4 which was the best model available at the time of the study.} chatbot into three CS-related courses offered at Aalto University. The integration with the LLM was unfiltered, meaning that students could also discuss contents unrelated to the courses. The closest match to our study was recently conducted by Prasad et al.~\cite{prasad2023generating}, who provided unrestricted access to an LLM through a programming environment plugin, and explored how students used the LLM. While a lot of concerns about potential student over-reliance have been raised in previous work~\cite{prather2023robots}, combined with the prior study by Prasad et al.~\cite{prasad2023generating}, our work provides further information on how students use an unrestrained LLM chatbot that is purposefully similar to commercially available LLM-based chatbots such as ChatGPT. 

Our research questions for the present study are as follows: (\textbf{RQ1}) How does the use of the LLM-based course assistant relate to the course?; (\textbf{RQ2}) How does the perceived usefulness of the LLM-based course assistant relate to the course?; and (\textbf{RQ3}) How does the use of the LLM-based course assistant relate to student background, and prior experience with LLMs?

\section{Background}

Computing education as a field has been continuously evolving, influenced by increases in processing power, availability of personal computers, access to the internet, online learning management systems, open online courses, and most recently large language models. The amount of learners is huge, bolstered by initiatives such as Computer Science for All~\cite{house2016fact} and Hour of Code~\cite{hourOfCodeStats}, where the latter has reported over a hundred million students. 

Programming education---a part of computing education---and the ways how programming education could be improved is a significant research topic in computing education research~\cite{luxton2018introductory}. When learning to program, students learn to understand both syntax and semantics of the programming language, slowly acquire plans that are used to reach reoccurring goals, and learn pragmatic aspects such as working with the available tools~\cite{du1986some}. This takes plenty of effort and can be challenging; nearly one-third of introductory programming students in higher education fail the introductory programming course~\cite{watson2014failure}. Improving pedagogy, classroom design, and instruction can help improve retention, although even after improvements, there still exists a considerable body of students who fail to succeed~\cite{vihavainen2014systematic}. 

Courses employ teaching assistants in a variety of tasks including assisting students in programming labs, grading assignments, and giving office hours~\cite{mirza2019undergraduate}. With infinite qualified teaching assistants, a classroom could in principle employ one-to-one mastery learning, which has been shown to lead to two standard deviation improvement in learning outcomes when compared to students in traditional classroom~\cite{bloom19842}. However, the increasing enrollments in programming classrooms and the associated costs would make this unfeasible. Thus, programming classrooms often use automated assessment systems~\cite{paiva2022automated}. Automated assessment systems can enhance the efficiency and scalability of the assessment process, making it possible to provide immediate feedback to students and ensure a fair and unbiased evaluation as every student's work is subjected to the same criteria. Despite the benefits, automated assessment has limitations as it mainly focuses on assessing the correctness of student-written code, failing to capture and aid in the problem-solving process~\cite{keuning2018systematic}. 

The recent rise of large language models has been highlighted as an additional avenue that could improve---and will certainly impact---computing education~\cite{becker2023generative,brusilovsky2023significant,denny2023computing}. Researchers have already explored the capabilities of large language models, highlighting their potential in writing code and solving and creating programming assignments~\cite{sarsa2022automatic, Finnie-Ansley2022Robots, puryear2022github}, explaining code~\cite{macneil2023experiences, leinonen2023comparing}, identifying programming concepts~\cite{tran2023using}, improving programming error messages~\cite{leinonen2023using, santos2023always}, and responding to students' help requests~\cite{hellas2023exploring}. 
While most of the studies on generative AI and LLMs in the context of computing education are based on expert evaluation of model outputs and single-shot experiments, the studies highlight the potential of using LLMs for formative feedback. 

To highlight this, studies exploring the use of LLMs as teaching assistants are starting to emerge~\cite{kumar2023impact,kumar2023quickta,prasad2023generating}. While expert evaluations and one-off experiments provide valuable insights, there is a need for further studies that consider the potential and impacts of LLM-based teaching assistants---or chatbots---in computing education. Perhaps the closest match to our work is that of Prasad et al.~\cite{prasad2023generating} who found that students did not use LLMs much beyond the assignment where they were introduced in an upper-level course when students were provided free, unrestricted access to LLMs through a visual studio code plugin.

\section{Methodology}

\subsection{Context and chatbot}

The experiments were conducted at courses offered by Aalto University in Finland. The courses in question use an online learning platform that allows hosting interactive ebooks with embedded assignments. During the summer of 2023, we integrating a LLM-based chatbot to the course platform. The chatbot was based on OpenAI APIs\footnote{We used the state-of-the-art model, GPT-4.} and students could engage in dialogue with it. We intentionally did not conduct any prompt engineering to e.g. constrain or modify the responses, and allowed students to use the chatbot similarly to how one would use a dialogue-based system like ChatGPT. 

The chatbot is available for learners on each course material page. When clicking an icon indicating the chatbot, the chatbot opens up in a modal window in which it can be conversed with. Students were made aware that any communication that they had with the chatbot would be stored both on the course platform and sent to the OpenAI APIs. The use of the chatbot was limited to 5 messages per minute and to 100 messages per day.

The platform and the courses that use the chatbot informed students of the chatbot and the policy associated to using the chatbot, highlighting that the chatbot is an assisting technology and that using the chatbot for creating solutions for assignments is not acceptable. The policy of use was provided as follows.

\begin{tcolorbox}[breakable]

In the Fall of 2023, we introduced a large language model -based generative AI assistant to the course platform. You can find it on the lower right corner of the material pages when logged in -- clicking it opens up a chat. The current version of the course assistant is based on ChatGPT. The assistant is not a TA, but a tool to help you with the course.

Similarly to asking information from your peers, course teachers, and TAs, you can use the AI assistant to help you with the course and the materials. You can, for example, ask for it to provide additional information about a topic, to explain code, to identify bugs in your code, and so on. You can also ask it for help when you are stuck e.g. with a programming assignment.

\begin{quote}
There are humans available for help as well, as discussed in the part on ``Asking for help and discussion area''.
\end{quote}

Do not use the assistant for creating solutions to the assignments, or ask it to complete the assignments for you, as this is harmful for learning. Like using solutions from others, using solutions generated by generative AI and large language models constitutes as plagiarism.

\begin{quote}
    The use of generative AI and large language models such as ChatGPT for completing coursework on your behalf is not allowed. Using solutions from ChatGPT or similar relates to representing the work of others as your own. When submitting coursework, only use solutions constructed by yourself.
\end{quote}

If you are uncertain whether your use of the assistant is allowed, please ask the course staff, and keep in mind that you are responsible for your own learning. A good way to rehearse and assess whether you have internalized the concepts and that you have worked on with your peers, TAs, or the assistant (etc), is to take a 30 minute break after the collaboration and complete (or redo) the problems on your own.
\end{tcolorbox}

\subsection{Courses}

During the fall of 2023, the chatbot was in three courses offered using the platform: (1) Software Engineering with Large Language Models, (2) Device-Agnostic Design, and (3) Web Software Development. The Software Engineering with Large Language Models (SE with LLMs) course was a tailored course for software engineers working in the industry. The course introduces principles of LLMs, including how they work and how they are prompted, and broadly discusses leveraging them in different phases of the software development life cycle. Students completed tasks from the software development life cycle, including documentation tasks, programming tasks, and testing tasks. 

The Device-Agnostic Design (DAD) course is a first-year MSc course that focuses on the principles of designing applications that work on a wide range of devices with multiple possibilities for input modalities. The course projects used Dart and Flutter as the technologies. Finally, the Web Software Development (WSD) course is a 2nd year Bachelor's level computer science course where students learn to design and implement web applications. In the course, students used Deno and Hono for building server-side functionality. Notably, the course tries to leverage new technologies and introduced also Deno KV\footnote{A globally replicated low-latency key-value database announced in May 2023.} and used Svelte and SvelteKit\footnote{The course used Svelte 5 alpha, which was released in November 2023, one week before the first lecture that focused on building client-side functionality.} for building the client-side functionality.

One of the authors of this article is the responsible teacher of all of the three courses. 

\subsection{Surveys and feedback}

The course platform had a brief background survey and a brief feedback item for providing feedback on the utility of the LLM based chatbot. Providing survey answers and feedback was voluntary and students were not compensated for answering in any way.

\subsubsection{Background survey}
\label{subsubsec:background-survey}

The background survey asked for experience in programming and in the use of LLMs. The questionnaire contained three items, which were as follows.

\begin{enumerate}
    \item On a range from `Not at all experienced' to `Very experienced', how would you characterize your prior programming experience?
    \item If you have written programs before, in lines of code, what is the largest program you have written? (NA=not applicable)
    \item On a range from `Not at all experienced' to `Very experienced', how would you characterize your experience of using large language models (e.g. ChatGPT, GitHub Copilot, ...)?
\end{enumerate}

Items 1 and 3 were answered using a scale from 1 to 9, where 1 corresponded to not at all experienced, while 9 corresponded to very experienced. Item 2 was responded to using the following options: NA, Under 500, 500-5000, 5001-40000, and over 40000.

\subsubsection{Usefulness of the chatbot}

At the end of every dialogue with the chatbot, the course material opened a dialog with the question ``How useful was the chatbot?'' that could be answered with a rating ranging from 1 star to 5 stars.

\subsection{Data collection and filtering}

All data was collected during the Fall of 2023 and processed accordingly to the national ethical guidelines. No ethical review was required. Overall, during the Fall of 2023, 257 students used the chatbot. From these, 228 provided research consent (89\%). From the 228, 14 did not actively participate in any of the courses and were omitted from the analysis. Thus, the analyses on chatbot usage focus on 214 students. As the background survey was optional, not all 214 students provided background data.

\subsection{Usage coefficient analysis}

In order to study whether there are differences between courses and course chapters in how students used the chatbot, we calculated a usage coefficient for each course chapter. First, we calculated the average chatbot use for each student separately. Then, for each student, we calculated a coefficient comparing their chatbot use in each chapter to their personal average. For example, a coefficient of 0.5 would indicate that the student used the chatbot only half of their usual average, while a coefficient of 2 would mean they used the chatbot twice as much as their average for a specific chapter. Then, for each chapter, we calculated the average coefficient over all students. This allows calculating statistics such as the ranges and the standard deviations of the coefficients per course, both of which can indicate the magnitude of differences between chapters. The results of this analysis are presented in Section~\ref{sec:chatbot-usage-per-chapter}.

\section{Results}

\subsection{Descriptive statistics}

Descriptive statistics of the chatbot usage per course are outlined in Table~\ref{tbl:overall-usage}.\footnote{Note that as a few students took part in multiple courses, the sum of the individual courses' participant counts is slightly larger than the total number of students overall.} The usage of the chatbot was highest in the SE with LLMs course, where 98\% of the participants used the chatbot. In the two other courses, the usage was lower, where 22\% and 24\% of the participants used the chatbot for the DAD course and the WSD course, respectively. As shown in the table, the amount of messages also differed considerably between the courses, ranging from an average of 4 messages per chatbot user (in DAD) to an average 100 messages per chatbot user in SE with LLMs.

\begin{table}
\caption{Descriptive statistics of the chatbot usage per course.}
\label{tbl:overall-usage}
\begin{tabular}{lr r }
\toprule
 Course & Users & Messages \\
 \midrule
(1) SE with LLMs                                    & 59 / 60  (98\%)    &  5916  \\
(2) DAD                          & 24 / 109  (22\%)   &  99   \\
(3) WSD                        & 135 / 554 (24\%)   &  1094 \\
\bottomrule
\end{tabular}
\end{table}

Table~\ref{tbl:prior-experience} outlines statistics from participants' self-reported prior experience collected using the survey outlined in~\ref{subsubsec:background-survey}. Overall, participants in SE with LLMs rated their programming experience as somewhat higher than those in the other courses. On the other hand, participants in the other courses self-reported their experience with LLMs as somewhat higher than participants in SE with LLMs. 

\begin{table}
\caption{Participants' self-reported prior overall programming experience (Prog. Exp.), programming experience in lines of code (LOC), and experience of using large language models (LLM exp.). Symbol $\mu$ denotes mean and symbol $\eta$ denotes median.}
\label{tbl:prior-experience}
\begin{tabular}{lccccccc}
\toprule
 Course    & n & \multicolumn{2}{c}{Prog. Exp.} & \multicolumn{2}{c}{LOC} & \multicolumn{2}{c}{LLM Exp.} \\
 \cmidrule(lr){3-4} \cmidrule(lr){5-6} \cmidrule(l){7-8}
  & & $\mu$ & $\eta$ & $\mu$ & $\eta$ & $\mu$ & $\eta$\\
 \midrule
(1) SE with LLMs  & 47 & 6.1 & 7 & 2.4 & 2 & 2.7 & 2 \\
(2) DAD           & 8  & 5.1 & 6 & 1.8 & 2 & 4.5 & 4  \\
(3) WSD           & 73 & 5.0 & 5 & 1.9 & 2 & 4.2 & 4  \\
\bottomrule
\end{tabular}
\end{table}

\subsection{Chatbot usage per course}

Chatbot usage was highly variable between the students, producing a distribution that, at least visually, appears roughly Zipfian. While the two most active users had over 400 messages with the chatbot, the third most active student had under 300 messages. Of the 214 students who used the chatbot, 18 students produced over one-half of the messages. Only 61 (28.5 \%) had 25 or more messages, while 85 students (39.7 \%) had 10 or more messages. 

\begin{figure}
    \centering
    \includegraphics[width=0.47\textwidth]{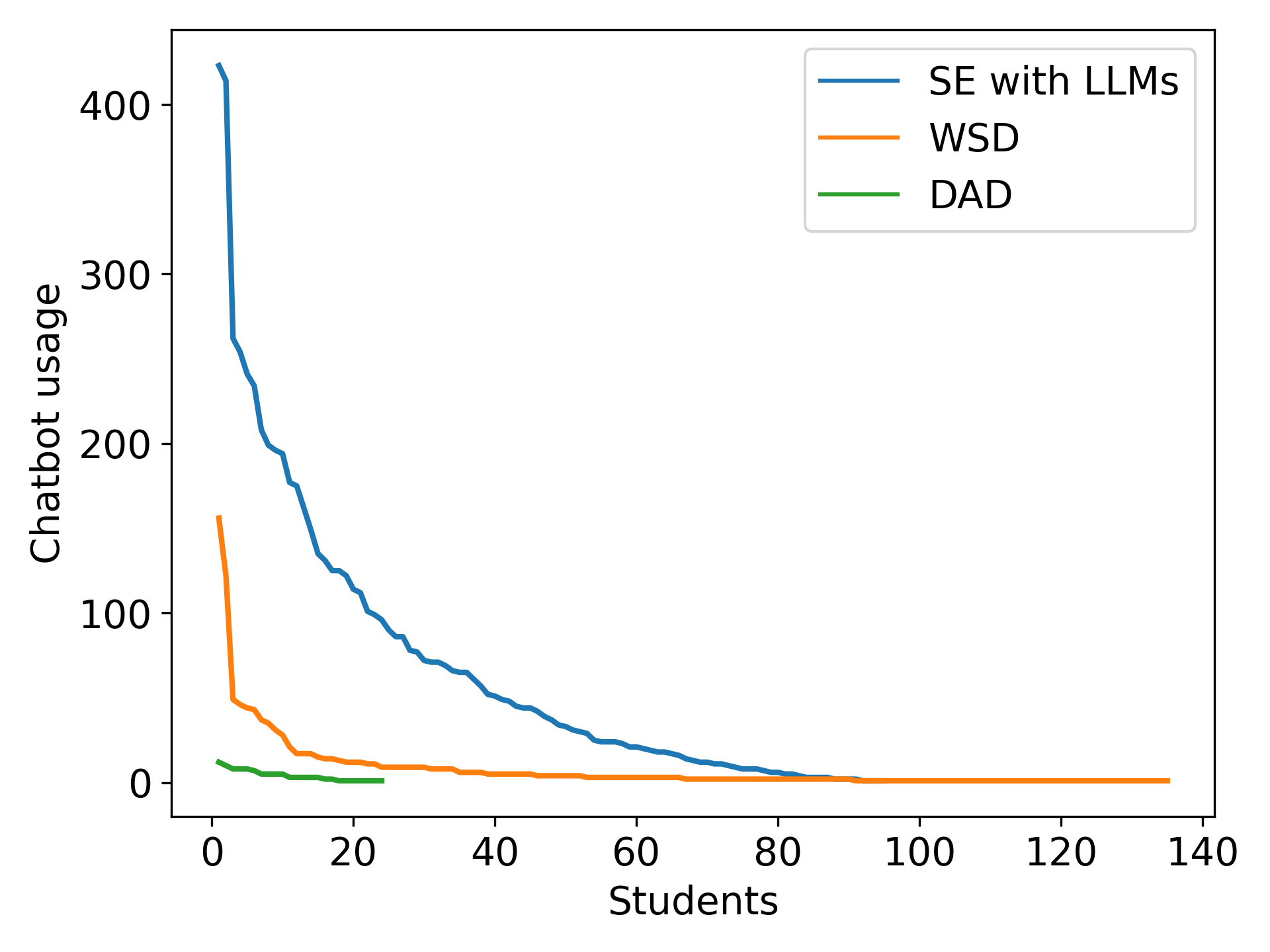}
    \caption{Student chatbot usage distribution per course.}
    \label{fig:per-course-usage-distribution}
\end{figure}

Observing the courses in isolation, we note that both the SE with LLMs course, and the WSD course, exhibit the same phenomenon. In both courses, two power users have significantly more interactions with the chatbot than others students, with the number of interactions quickly decreasing. These top power users have 423 and 414 messages in the case of SE with LLMs, and 156 and 122 interactions in the case of WSD. For comparison, the third-most active users on these courses have 262 and 49 interactions, respectively. These power users are distinct students. For the Device Agnostic Design course, the usage levels are generally very low, with the highest number of messages for any student being 12. The per-course usage distributions are shown in Figure~\ref{fig:per-course-usage-distribution}.

\subsection{Chatbot usage per chapter}
\label{sec:chatbot-usage-per-chapter}

\begin{table}[t]
    \caption{Chatbot Usage Coefficients per Course}
    \centering
    \begin{tabular}{lcccc}
    \toprule
        Course & Mean & Median & SD & Range \\
        \midrule
         (1) SE with LLMs & 0.91 & 0.97 & 0.45 & [0.26, 1.77] \\
         (2) DAD & 1.00 & 0.97 & 0.23 & [0.72, 1.32] \\
         (3) WSD & 0.98 & 0.96 & 0.22 & [0.64, 1.45] \\
         \bottomrule
    \end{tabular}
    \label{tbl:usage-per-chapter}
\end{table}

We further looked into the use of the chatbot in individual chapters of the material, focusing on deviations from average usage behavior. The aggregate statistics for the three courses are shown in Table~\ref{tbl:usage-per-chapter}. From the table, we can see that usage was quite similar between the WSD course and the DAD course, but the SE with LLMs course showed different usage behavior. In the SE with LLMs course, the standard deviation and the range of coefficients was larger, suggesting that there were larger differences between individual chapters of the material in how much students used the chatbot.

Overall, in the SE with LLMs, the chatbot was most used in a chapter on tooling and working with Python, which included a range of programming problems, explicitly allowing students to use the chatbot for solving them to demonstrate LLM code generation capabilities. The second chapter with the most use was a chapter that focused on building a larger application with the help of LLMs, starting from decomposing the problem and resulting with an application with a graphical user interface. The chapters with the least chatbot usage focused on review and testing and software engineering, neither of which had assignments and both of which discussed the topics on a higher level.

In the DAD course, the least usage was observed in a chapter on Flutter basics, which introduced participants to showing simple content in a Flutter application, while the most usage was observed in a chapter on handling input with Flutter. Both chapters included programming problems, but the problems in the chapter on handling input were considerably more complex 

In the WSD course, the chatbot was most used in a chapter introducing the concept of storing data on server using Deno KV and a chapter on state management with Svelte. Notably, in both of these chapters, the usefulness feedback median was 1, indicating that the chatbot was not at all useful -- very likely due to the technologies being so new that the LLM would suggest deprecated approaches. The chatbot was least used in a chapter on data validation, which introduced the principles of validating data, and introduced a library for the task. 

\subsection{Chatbot usefulness}

At the end of every dialogue, students were prompted to rate the usefulness of the chatbot using a rating from 1 to 5 stars. Table~\ref{tbl:chatbot-usefulness} outlines the results. Overall, students in the SE with LLMs considered the chatbot as somewhat more useful (avg. rating 3.8/5), than students in the other courses who rated the chatbot on average 3.1/5. The median usefulness in all of the courses was 4 out of 5. These numbers, however, need to be considered in the context of a possible self-selection bias: a student who trials the chatbot once and determines it unusable would produce only a single low rating, while superusers happy with the chatbot might produce hundreds of high ratings. We return to this topic in the discussion.

\begin{table}
\caption{Average usefulness of the chatbot in each of the courses. The median usefulness was 4 (out of 5) in all courses.}
\label{tbl:chatbot-usefulness}
\begin{tabular}{lr c }
\toprule
 Course    & Ratings & Average Usefulness \\
 \midrule
(1) SE with LLMs   & 456  &  3.8 \\
(2) DAD            & 35   &  3.1 \\
(3) WSD            & 242  &  3.1\\
\bottomrule
\end{tabular}
\end{table}

\subsection{Usage and student backgrounds}

Using Spearman's Rho (Table~\ref{tbl:correlations}), we also observed that chatbot usage, measured as the number of messages, was moderately negatively correlated with students' previous experience with LLMs ($\rho=-0.41, p < 0.01$). We hypothesize that this decrease in usage with experience can be explained by inexperienced students running various tests and trials to get a better feel of the LLM, which those already familiar with LLMs would presumably not conduct at least to the same degree. On the other hand, as we briefly discuss in our study limitations, students with more experience with LLMs may have access to LLMs through other means. The correlations between chatbot usage and prior programming experience or largest program written were not statistically significant ($\rho=0.10$, $p=0.25$ and $\rho=0.11$, $p=0.20$, respectively).

\begin{table}
\caption{Correlation coefficients between chatbot usage and student experience variables. Data contains only those students who responded to the experience survey. Bolded values have $p < 0.001$.}
\label{tbl:correlations}
\begin{tabular}{lrrrr}
\toprule
 & Usage    & Prog. Exp. & LOC & LLM Exp. \\
 \midrule
 Usage        & \textbf{1.00} & 0.10 & 0.11  & \textbf{-0.41} \\
 Prog. Exp.   &    & \textbf{1.00}   & \textbf{0.58} & 0.16 \\
 LOC          &    &      & \textbf{1.00}    & 0.19 \\
 LLM Exp.     &    &      &       & \textbf{1.00} \\
 \bottomrule
 \end{tabular}
 \end{table}

\section{Discussion}

\subsection{Course and population differences}

The courses differed in terms of participants and chatbot usage. The SE with LLMs was attended by software engineers from the industry who rated their prior programming experience higher than students in the other courses. At the same time, students rated their prior experience with LLMs higher than the software engineers. The differences in programming experience was to be expected, while we were somewhat surprised with the difference in LLM experience. Students may be more active in looking for help from new sources and more likely to adapt new tools as they come; indeed, in our context, students are actively discussing LLMs, which is also visible in the quantity of theses related to LLMs. Moreover, software engineers might be constrained in terms of the tools that they can adapt, and larger companies can still be vary of LLMs due to existing legal disputes\footnote{See e.g. \url{https://githubcopilotlitigation.com/}} and uncertainties.

\subsection{Chatbot usefulness}

Our results indicate that at least \textit{some} students find LLMs highly useful, becoming powerusers, but at the same time a significant amount of students barely engage with them. As the usefulness ratings collected from the students are dominated by the first group, they should be interpreted with caution: while the intrinsic component of the evaluation was positive, it suffers from a potential self-selection bias and the main proxy for extrinsic effectiveness, actual usage, offers a less clear view. Further study is clearly needed. 

At the same time, these results on the use and usefulness of LLMs were also to be expected. While the SE with LLMs course explicitly instructed participants to use LLMs, the other courses offered the chatbot more as an additional support mechanism. This already can impact the use of the chatbot significantly. In our case, less than 10\% of the students who used the chatbot produced more than 50\% of the messages. This also aligns with prior work that found that most students did not use an LLM-based chatbot beyond initially trying it out when it was introduced~\cite{prasad2023generating}.

When considering the relative differences in how students used the chatbot in the courses, we see parallels to the use of help resources in online courses. Students differ in how, when, and from whom they ask for help~\cite{nelimarkka2018social}; in online courses, the majority of participants do not engage in discussions, while some are very active, even to be labeled as ``superposters''~\cite{nelimarkka2018social,huang2014superposter}. As prior research has highlighted that there are students who perhaps read posts but do not necessarily comment on posts or ask questions~\cite{kizilcec2017self}, a possible future stream of research would be to identify and highlight discussions with the chatbot that have been very useful for learning, and allow sharing them to other course participants on the online platform.

Similarly, the average usefulness differed between the courses. The higher usefulness of the chatbot likely relates to the direct use for course tasks, while the lower usefulness in other courses could relate to course technologies. Both DAD and WSD have students work on larger projects with multiple files, which might not be very convenient with the chatbot. In addition, both courses also keep up to date with technology versions; as an example, WSD used Svelte 5 alpha as the frontend technology, which was released just before the start of the classes that focused on building frontend functionality. As LLMs have a knowledge cutoff point that reflects the time when the data that was used for training, LLMs in general do not have information of technologies released after specific moments in time. One potential direction for future work would be to add retrieval augmented generation functionality to the course materials, which would allow the LLMs to retrieve information from the materials when creating a response.

\subsection{Instructor viewpoint}

Overall, when considering the possibility of embedding an LLM-based chatbot to the course platform, we noted that students are already gaining experience from using LLMs and some students also have accounts to services such as OpenAI ChatGPT. By providing access to a chatbot that leverages a state-of-the-art LLM, we created a possibility of leveling the playing field, where students could use a state-of-the-art LLM even if they would not be paying for it. The cost of using the OpenAI API was less than \$200 for the whole fall semester. 

While the SE with LLMs was a new course, DAD and WSD are courses that have been offered in previous years. The courses allow students to ask for help through the online platform and offer labs for students where they can ask for help. The introduction of the LLM-based chatbot did not affect the use of the help request functionality or the labs in an observable manner. Prior research has pointed out that making sample solutions available to students can lead to reduced use of support~\cite{nygren2019non}; although the study contexts are different, we can speculate that the LLM-based chatbot was not simply seen as a source for sample solutions.

The courses also use tools for plagiarism detection that use fingerprinting and compare solutions to detect similarities. We did not observe noticeable differences in plagiarism, nor did we identify students explicitly seeking to simply use the models for solving their assignments. We however acknowledge that detecting LLM-generated content can be difficult, and the courses may have suffered from the problem already earlier. 

Perhaps one of the key observations from class discussions was that the chatbot was rather poor at helping with errors and with debugging larger code, which was also included in the few written feedbacks. While earlier research has highlighted the possibility of using LLMs for enhancing programming error messages~\cite{leinonen2023using, santos2023always}, the prior studies have been conducted with introductory-level programming assignments that are relatively small and well-scoped. In our context, the applications where students needed help were typically larger, consisting of multiple files. This highlights also the need to consider moving towards IDE-integrations in chatbots, as has already been done with e.g. GitHub Copilot.

\subsection{Limitations of work}

This study comes with a range of limitations, which we discuss here. First, we acknowledge that students may have used LLMs also through other means, for example through their own accounts on LLM providers. While we did not ask whether students used other LLMs, it is possible that this could be the case. Second, responding to the various survey instruments was voluntary. While we received over 700 responses to the brief usefulness rating that students could answer by clicking once, as the feedback was tied to actually using the chatbot, the responses naturally underrepresent students who did not extensively---or at all---use the chatbot. Third, as the student background was self-reported, it might not accurately reflect the real ground-truth experience levels of the students. Fourth, while we built the chatbot on the OpenAI APIs, we did not collect information on how much time creating the response took. Especially on larger responses, the time to produce the response can be considerable, which by itself can already reduce the perceived usefulness of the chatbot. 

We also acknowledge that OpenAI APIs were under a denial of service attack on a few days, which could also influence the time to form a response -- or even whether a response was formed at all. We also note that the chatbot was embedded into the learning materials, and not e.g. to any programming environment that the students used. This likely influences the usefulness of the chatbot especially with larger assignments. Finally, this type of ``intrinsic'' evaluation of a natural language generation system is generally viewed as inferior to an ``extrinsic'' evaluation focused on how the system allows the user to complete some specific task~\cite{celikyilmaz2020evaluation,van2021human}. Our analysis also focuses on the \textit{average} performance of the chatbot, an approach that ignores the potential for catastrophic errors that---even if rare---could meaningfully affect whether a chatbot like this is in reality an ethically feasible learning aid~\cite{novikova-etal-2017-need}.

\section{Conclusion}

In this work, we report our experiences from giving students access to a LLM-based chatbot. We purposefully made the chatbot similar to commercially available chatbots, so that student use would hopefully be as authentic as possible. Our experiences suggest that the worst fears of educators---most students developing severe overreliance on LLM chatbots---might not materialize, even if the chatbot does not have any ``guardrails''~\cite{liffiton2023codehelp}.

To summarize, our answers to the research questions are as follows: (RQ1) The use of the chatbot differed considerably between the courses and to some extent between course material chapters, where the chatbot was most used in a course that taught participants to use LLMs in software engineering. The other two courses saw less use of the chatbot, and there was less variation in the amount of use between course materiel chapters. (RQ2) The chatbot was perceived as most useful in the course that focused on LLMs in software engineering (average 3.8/5), while participants in other courses rated the interaction as somewhat less useful (average 3.1/5). In all courses, the feedback median for usefulness was 4. Anecdotal evidence highlighted that the students did not find the chatbot very useful for debugging or improving error messages, despite some previous works highlighting LLMs' potential for these tasks~\cite{santos2023always,leinonen2023using}. Similarly, the usefulness of the chatbot in course material chapters that involved very recent technologies was low. (RQ3) Previous experience with LLMs was linked with lower use of the chatbot, but prior programming experience was not related to the chatbot usage. 

As a part of our future work, we are exploring the relationship of self-regulation, large language model use, and perceptions and conceptions of plagiarism.

\begin{acks}
This research was supported by the Research Council of Finland (Academy Research Fellow grant number 356114).
\end{acks}


\balance
\bibliographystyle{ACM-Reference-Format}
\bibliography{sample-base,99-sori-references,99-refs}

\end{document}